# Analysis of Requirements for the Design of a Detector Control System in a High Energy Physics (HEP) Experiment

**Juan Carlos Cabanillas Noris[1]**

*Technological Institute of Culiacan*
*Juan de Dios Bátiz Avenue 310, 80220 Culiacán, México*
*E-mail:* `jccabanillas.noris@itculiacan.edu.mx`

**Ildefonso León Monzón, Solangel Rojas Torres**

*Autonomous University of Sinaloa*
*E-mail:* `ildefonso.leon.monzon@cern.ch`

**Mario Iván Martínez Hernández**

*Autonomous University of Puebla*
*E-mail:* `mim@fcfm.buap.mx`

In this work the use of the Rational Unified Process (RUP) to model the design of a Detector Control System (DCS) in a High-Energy Physics (HEP) experiment is proposed. We include a brief description of the wide diversity of elements and aspects to be considered and general definitions for the analysis of requirements. The characteristics, functional and non-functional requirements and Use Cases (UC) of main actors involved in the design, implementation and operation of this type of software systems are also specified. For the description of these requirements, tools of the Unified Modeling Language (UML) like diagrams and tabular use cases are used by means of the Enterprise Architect® software.

---

[1]Speaker

1. Introduction

Precise knowledge of the requirements of a system is a principle that unquestionably marks the success or failure of any software development related to that system. Establishing and managing the requirements for the design of a system are fundamental aspects of software engineering. One of the main software systems in a High Energy Physics (HEP) experiment is the Detector Control System (DCS). The DCS allows the control, monitoring and configuration of the experimental equipment either from a single on-site place or remotely, through a Supervisory Control and Data Acquisition (SCADA) system. The DCS manages communications and provide feedback to the various experiment's services and systems, such as: the accelerator, cooling systems, electricity, magnets, safety, access control, among others.; providing feedback to them.

On the other hand, the design, simulation, construction, installation, commissioning and operation of a detector in a HEP experiment are usually complex tasks. In addition to the former, it is necessary to consider the large number of detectors and subdetectors that need to be controlled independently by corresponding DCS, there are significant conceptual and structural differences among them, such as: types of events to be analyzed (particles collisions or cosmic rays), dimensions, materials (semiconductors, plastic scintillators, gas chambers), parameters to be monitored (temperature, radiation levels, magnetic field, position, cryogenic system, etc.), type of control actuators (cooling, high and low voltage channels, gas system, etc.), infrastructure, number of readout channels, etc. [1]. This complexity and diversity of detectors should always be considered when defining and specifying the requirements of the experiment's control system. Based on the previous description, an initial methodology that specifies and standardizes the requirements for the development, commissioning and operation of the software for a HEP experiment's DCS was carried out.

2. Requirement Analysis Overview

Modeling consists in creating representations of system abstract models, where a diagram or table represents a different perspective of it. Models permit: 1) to know system requirements (in this case a software for the control of a detector for HEP studies), and 2) to generate a system abstraction to simplify and compiling its most important characteristics [2].

Firstly, the model process named Rational Unified Process (RUP), which is a case-driven language, architecture-centric, iterative and incremental. Unified Process is repeated over a series of cycles, where each cycle constitutes a version of the system [3]. A cycle in RUP consists of four discrete phases in the software processes, which are: conception, elaboration, construction and transition. Each phase can be presented iteratively with the incremental delivery of results. An iteration is a set of disciplines or workflows are developed, among which the following stand out: requirements, analysis, design, coding, and testing. Similarly, in RUP, description of the workflow is oriented on associated UML models, such as: use case, sequence, analysis, design, etc. [3].

This work focuses on requirements specification of the RUP model process, which is are useful for documentation of the software's required performance from the point of view of the main actors involved. It generally uses a mathematical notation or modeling language [3]. For that a software engineering analysis technique called *use cases* has been used. In general, a use case shows a generic scenario describing what a particular user requires from the system [3].

Use-case modeling should preferably include a diagram and a tabular description. In the following sections, general characteristics, functional and non-functional requirements, and elements of the use cases for the DCS of a generic HEP experiment are described

**2.1 Actors (stakeholders)**

Requirements analysis of the control system is done based on the three main actors that, in our case, are:
1. Expert in the Detector (ED)
2. Expert in the Central DCS of the Experiment (ECDCS)
3. Operator in the Central DCS / DCS Shifter of the Experiment (OCDCS)

An actor can be a person or a subsystem (software, modules, logbook, web browser, etc.).

**2.2 Objetive**

"Develop a methodological proposal for obtaining software requirements related to design, commissioning and operation processes of the detector control system (DCS) for High-Energy Physics (HEP) experiments applying software engineering techniques".

**2.3 Justification**

- Upgrading existing detectors or installing of new detectors in a High Energy Physics (HEP) experiment implies the development, integration and commissioning of the control system for each detector in the central DCS.
- The definition of a methodology for the development process of the DCS for a new or updated detector applying software engineering techniques has the following advantages:
  a) The stakeholders involved in the process of control software development, as well as the description of their general characteristics are clearly defined..
  b) The functional and non-functional requirements by applying diagrams and tables of use cases; in addition to the specification of these by means of activity diagrams.
  c) The processes of static and dynamic structures, where the main stakeholders are involved, are defined in a standardized way.
  d) The behavior of the nodes of the software architecture for a detector, specifying the sequences of states and their response to certain events, usually applying Finite State Machines (FSM).

**2.4 Tablular description of use cases**

A tabular description shows detailed information about the specific use cases appearing in a diagram. These tables include the following elements:
- *Actors:* name of those involved in the use case.
- *Description:* brief explanation of what constitutes the use case, and how actors will use required data.
- *Data:* information actors require to carry out a process.
- *Stimulus:* internal or external action that triggers the start of a process.
- *Response:* defines how the system should respond to a stimulus according to provided information.
- *Comments:* annotations and important observations concerning the particularities of each case and/or involved elements.

3.     Requirements of the Expert in the Detector

**3.1 General Characteristics (ED)**

- Involves in the design, commissioning, maintenance and updating of the control system of the detector.
- Restores the normal functioning of the detector in case of any disturbance or unwanted operating conditions.
- Attends the queries or requirements of the experiment's operators on shift.
- Implements the recommendations and/or updates to the system from the staff of the central DCS.

**3.2 Functional Requirements (FR-ED)**

In the table I, use-cases of the software functional requirements by expert in the detector (ED) are shown.

Table I. Use Cases Diagram of Functional Requirements of the ED

| RF Code | Functional Requirements of the Expert in the Detector |
|---|---|
| UC_FR01_01 | Defining the Interaction Using User Interfaces |
| UC_FR01_02 | Set up Alarms |
| UC_FR01_03 | Set up Alarm Help Instructions |
| UC_FR01_04 | Plotting the Behavior of Important Parameters |
| UC_FR01_05 | Model the Behavior of the DCS Nodes through FSMs |
| UC_FR01_06 | Design DCS to Frond-End Electronic Link |
| UC_FR01_07 | Integrate the DCS Detector to Central DCS |
| UC_FR01_08 | Configuring of Flows Parameters that Determinate Operating Conditions |
| UC_FR01_09 | Configuring of Restricted Access Based on Roles |
| UC_FR01_10 | Operate and Monitor the Power System |
| UC_FR01_11 | Configuring of Runs Parameters |

For each of the functional software requirements of all the actors, their corresponding tabular use cases and diagrams were defined.

4.     Requirements of the Expert in the Central DCS (ECDCS)

**4.1 General Characteristics (ECDCS)**

- Supervises that systems and subsystems of the DCS of all detectors, which integrate the experiment, work correctly.
- Remains on alert in case of any disturbance or unwanted conditions in the central DCS, or when the operating personnel in turn of the control center require it.
- Trains to collaboration members who wish to participate as central operators of the DCS in the Run Control Center (RCC) of the experiment.
- Provides monitoring, training and advice in the processes of software migration and updating of the control systems for all detectors in the experiment.

**4.2 Functional Requirements (FR-ECDCS)**

In the figure 1, use cases diagrams of software functional requirements of the expert (ECDCS) and operator (OCDCS) in the central DCS are shown.

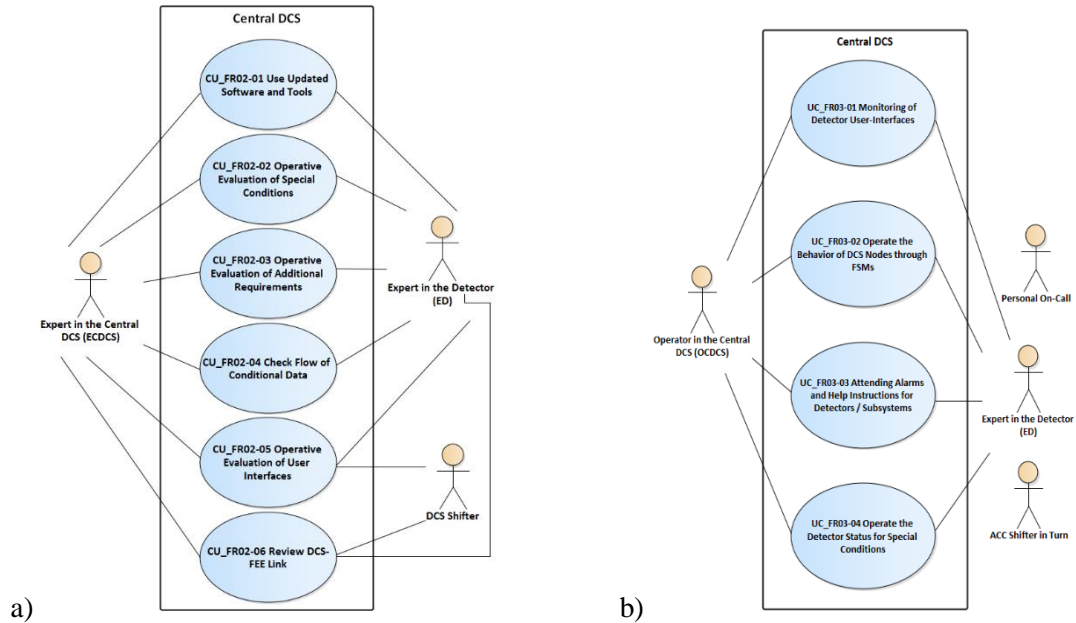

Figure 1. Use Cases Diagram of Functional Requirements of the a) ECDCS) and b) OCDCS

## 5. Requirements of the Operator in the Central DCS (OCDCS)

**5.1 General Characteristics (OCDCS)**

- Operates and monitoring the control system (DCS) of the experiment, and all its detectors and subsystems during a shift in the RCC of the experiment. Applies basic tools such as: a) User Interface (UI), b) Alerts and Errors Screen (AES), c) Operator and Experiment Logbook, and d) a repository of documents.
- Checks alert messages with the highest severity level on the AES and respond them with aim of restoring the normal operation of that subdetector or subsystem.
- Sets up the relevant safety conditions in the experiment for a set of detectors or a single detector using a special panel.
- Modifies the detector subsystems state when a change in the magnet (dipole/solenoid) conditions must be established, for example: planned ramp up/down transition of the current.
- Monitors and publishes important announcements in the DCS Main Screen at RCC.
- Could be assigned as safety leader during a shift, which is responsible of continuously monitoring screens related to security information, among other related activities.
- Develops a report of activities and incidents at end of the shift.
- Enables or disables the DCS locks (usually FSM) to take the control of a detector, sub-tree or node, after a shift leader request.
- Includes or excludes partially or totally one or more experiment detectors or subsystems.